\documentstyle[prb,preprint,aps]{revtex}
\begin{document}
\preprint{{\bf ETH-TH/95-21}}
\title{Vortex Dynamics and the Hall-Anomaly:
	\break a Microscopic Analysis}
\author{Anne van Otterlo$^1$, Mikhail Feigel'man$^2$, Vadim
	Geshkenbein$^{1,2}$, and Gianni Blatter$^1$}
\address{$^1$ Theoretische Physik, ETH-H\"{o}nggerberg,
	CH-8093 Z\"{u}rich, Switzerland\\
	$^2$ Landau Institute for Theoretical Physics,
	117940 Moscow, Russia}
\maketitle

\begin{abstract}
We present a microscopic derivation of the equation of motion for
a vortex in a superconductor. A coherent view on vortex dynamics
is obtained, in which {\it both} hydrodynamics {\it and} the vortex core
contribute to the forces acting on a vortex. The competition between
these two provides an interpretation of the observed sign change
in the Hall angle in superconductors with mean free path $l$ of the
order of the coherence length $\xi$ in terms of broken particle-hole
symmetry, which is related to details of the microscopic mechanism of
superconductivity.
\end{abstract}

\pacs{PACS numbers: 74.20.-z, 74.25.Fy, 74.60.-w}

In recent years the interest in vortex motion in superconductors has
revived, mainly due to the advent of high temperature superconductors
(HTSC). As a consequence of the peculiar material properties, the
physics of vortices in HTSC shows many new aspects not encountered
in conventional superconductors \cite{rmp}. A major theme is the sign
change in the Hall effect in the superconducting state, as is observed
in both HTSC \cite{exp,hagen} and conventional superconductors
\cite{hagen} for temperatures $T$ just below $T_{C}$.
This Hall-anomaly cannot be understood within the framework of the
conventional Bardeen-Stephen \cite{bs} or Nozieres-Vinen \cite{nv}
theories for vortex motion, that predict the Hall effect in the
superconducting and normal state to have the same sign for all
temperatures. Several attempts at a theoretical understanding of
the phenomenon have been undertaken \cite{tdgl,new}, but none of
these seem to explain the experimental data.
In Ref. \cite{hagen} Hagen {\it et al.}, comparing a number of experiments,
conclude that the sign change of the Hall effect is an intrinsic
vortex property that occurs if the electron mean free path $l$ is of
the order of the coherence length $\xi$. Within a phenomenological
analysis, Feigel'man {\it et al.}\cite{fglv} interpret the sign change
in terms of broken
particle-hole symmetry and obtain good agreement with the experimental
signatures\cite{hagen} of this effect. It is the purpose of this Letter
to report on a microscopic calculation of the dynamical single vortex
properties that yields a unifying description of the physics involved
and puts the results of the analysis of Ref.~\cite{fglv} on a firm
theoretical basis.

Before presenting the microscopic theory, we discuss our main results
for the vortex equation of motion and the resulting Hall force and
angle. In general one expects the forces on a vortex to consist of
two contributions, {\it i.~e.}, one from the electronic states in the
vortex core and one from the hydrodynamic flow far away from the core.
The vortex equation of motion has the form
\begin{eqnarray*}
	[M_{C}+M_{H}]\dot{{\bf V}}+\eta_{C}{\bf V}=
	\{\kappa_{H}{\bf V}_{T}-[\gamma_{C}+\gamma_{H}]{\bf V}\}
	\times {\bf z} \; .
\end{eqnarray*}
Here ${\bf V}$ denotes the velocity of the vortex and ${\bf V}_{T}$
is the transport velocity due to an applied current density ${\bf j}=
\kappa_{H}{\bf V}_{T}/\Phi_{0}$ [we consider a film or layered
structure, the extension to a 3D geometry is
straightforward]. The equation
of motion includes a vortex mass $M$, a damping term $\eta$, and the
Lorentz and Hall force coefficients $\kappa$ and $\gamma$. We made a
clear separation into core and hydrodynamic contributions by writing
subscripts $C$ and $H$ respectively. Extrinsic forces due to pinning
and the interaction with other vortices add to the r.h.s.\ of the
above equation of motion, however, here we consider only intrinsic
vortex properties. The coefficient for the Lorentz force,
$\kappa_{H}=\pi n_{s}$, arises from the hydrodynamic flow around the
vortex with the superfluid density described by $n_{s}$ \cite{sonin}.
The mass terms were considered by Suhl in a Ginzburg-Landau approach
\cite{suhl}. The core contributions $\eta_{C}$ and $\gamma_{C}$ were
calculated by Kopnin {\it et al.}\cite{kks}, who found [we use
$\hbar=c=k_{B}=1$]
\begin{equation}
	\eta_{C}=\pi n_{e}\frac{\omega_{0}\tau_{r}}
	{1+(\omega_{0}\tau_{r})^{2}}\; ;
	\;\; \gamma_{C}=\pi n_{e}\frac{(\omega_{0}\tau_{r})^{2}}
	{1+(\omega_{0}\tau_{r})^{2}} \; ,
\label{class}
\end{equation}
where $n_{e}$ denotes the electronic density,
$\omega_{0} \approx \Delta^2/\epsilon_{F} $ is the level
spacing between the localized Caroli-de Gennes-Matricon (CdGM) states
in the core \cite{cgm}, and $\tau_{r}$ is the relaxation time.

The key point in the determination of the Hall angle $\alpha_{{\rm Hall}}$
($\tan\alpha_{{\rm Hall}}=\gamma/\eta$) is to find the hydrodynamic
contribution $\gamma_{H}$. For comparison we remind the similar
procedure for uncharged Bosons like $^4$He, where the vortex core has no
internal structure and the Hall force arises from the first order time
derivative $\bar{\psi}i\partial_{t}\psi$ in the Lagrangian
density \cite{mpaf}. With $\psi=\sqrt{n}e^{i\varphi}$, $n$ the mean
particle density, the corresponding contribution to the Lagrangian
is $\delta {\cal L}= -n\dot{\varphi}$. In the presence of a vortex
at ${\bf R}$, the phase configuration is $\varphi({\bf r},\tau)=
\varphi_{v}({\bf r}-{\bf R}(\tau))$ with $\varphi_{v}({\bf r})=
\arctan(y/x)$. The Euler-Lagrange equation yields a Hall
force ${\bf F}_{{\rm Hall}}= - 2\pi n{\bf V}\times{\bf z}$, or
$\gamma_{H}=2\pi n$ for Bosons. If no normal fluid component is present
at $T=0$ the Hall and Lorentz force combine into the Galilei invariant
Magnus force ${\bf F}_{M}= \kappa
({\bf V}_{T}-{\bf V})\times{\bf z}$ \cite{sonin}.

A hydrodynamic contribution to the Hall force in a superconductor
arises also from a first order time derivative in the Lagrangian.
This is most clearly seen in a time dependent Ginzburg Landau (TDGL)
approach, where a term $\delta{\cal L}=(N_{e}'/2\Lambda N_{e})
\bar{\Delta}i\partial_{t}\Delta$ appears in the Lagrangian density
\cite{fu}. This term depends on the electronic band structure through
the derivative of the density of states $N_{e}$ at the Fermi level
$N_{e}'=\partial_{\mu}N_{e}(\mu)\mid_{\mu=\epsilon_{F}}$ and is thus
related to particle-hole asymmetry. Here $\Lambda$ is the strength
of the attractive BCS model interaction. Note that in BCS theory
$2N_{e}'/(\Lambda N_{e}) = N_{e}\partial_{\mu}\ln T_{C}$.
The same procedure as for $^4$He leads to a small hydrodynamic
contribution $\gamma_{H}= \pi (N_{e}'/\Lambda N_{e})
|\Delta|^{2}$ of order $n_{e}(\Delta/\epsilon_{F})^{2}$
($N_{e}'\simeq n_{e}/\epsilon^{2}_{F}$). Its exact magnitude and sign
depends on the (experimentally accessible) details of the electronic band
structure. Although core physics is lacking in a TDGL approach
\cite{tdgl}, we will see in the following that TDGL does predict the
correct hydrodynamic contribution $\gamma_{H}$. A hydrodynamic
contribution to $\gamma$ is a general property of superconductors
with broken particle-hole symmetry, also for temperatures far
below $T_{C}$.

Defining $n_{\Delta}=N_{e}'\Delta^{2}/(\Lambda N_{e})$, the total
Hall force constant for a superconductor at $\omega_0 \ll T\ll\Delta$
becomes (see also the detailed discussion in Ref.~\cite{fglv})
\begin{equation}
	\gamma=\pi n_{e}\frac{(\omega_{0}\tau_{r})^{2}}
	{1+(\omega_{0}\tau_{r})^{2}}+\pi n_{\Delta} \; .
\label{hall}
\end{equation}
The first term describes the contribution arising from the quasiparticles
bound to the core. Close to the superconducting-normal transition the
scattering states have to be included\cite{koplop} and the term
crosses over to the normal state Hall term; it therefore has
the normal state sign. The second term is the hydrodynamic contribution.
With $\omega_{0} = \Delta^2/\epsilon_F $ and
$n_{\Delta}/n_e \approx \Delta^2/\epsilon_F^2 $, the core term is
dominant in the clean limit $l > \xi(T)$, whereas the hydrodynamic
contribution determines the Hall angle in the dirty case. A sign change
in the Hall effect occurs if the hydrodynamic term has a negative sign,
{\it i.~e.}, $N'_e < 0$. Within a free-electron based BCS theory
we have $N'_e \geq 0$ and no sign change occurs. However, a simple
modification of the electronic dispersion can drive $N'_e$ negative,
resulting in sign changes of the Hall effect as described below
({\it e.~g.}, consider the dispersion
$\epsilon_{k}=k^{2}/2m+k^{4}/4m^{2}\epsilon_{0}$ in two dimensions:
The corresponding density of states is
$N_{e}(\epsilon)=m/\pi(1+2\epsilon/\epsilon_{0})$ and
$N'_e=-(2\pi/m\epsilon_{0})N^{2}_{e}(\epsilon)<0$ accordingly).
With $N'_e < 0$ the Hall effect has the normal state sign in the clean
limit and the opposite one in the dirty limit. Furthermore, the two
contributions have a different temperature dependence through
$\Delta(T)$, allowing for multiple sign changes.
Our interpretation of the sign changes in HTSC (Bi- and Tl-based
compounds) is as follows: at low temperatures the clean limit is
realized with $l > \xi(T)$ and the Hall effect has the normal state sign;
with increasing
temperature, $\tau_r$ and $\Delta$ decrease until the second term in
Eq.~(\ref{hall}) dominates and a first sign change occurs when $l \sim
\xi(T)$. At even higher
temperatures, close to $T_{C}$, the normal quasiparticles take over and
a second sign change back to the normal state sign occurs.
Note that the low-temperature sign change may be invisible if pinning
is strong enough, which is probably the case for YBCO. This
analysis provides a natural interpretation for the experimental findings
as summarized by Hagen {\it et al.}\cite{hagen}.

We now continue with an outline of the microscopic derivation of
the vortex equation of motion, starting from a model
Hamiltonian $H$ that includes a short range attractive BCS
interaction, as well as a long range repulsive Coulomb interaction
(see Ref.~\cite{futur} for more details).
We express the grand canonical partition
function as an imaginary time path integral over the electronic
fields $\psi$ and the gauge field $A_{\alpha}$ ($\alpha = \tau,x,y,z$),
\begin{equation}
	Z=\int{\cal D}^{2}\psi{\cal D}A_{\alpha} \exp\{-S\}
\end{equation}
with Euclidean action
\begin{eqnarray}\nonumber
	S= \int d^{3}{\bf r}\int^{\beta}_{0}d\tau {\Big (}
	\bar{\psi}_{\sigma}
	[\partial_{\tau}-ieA_{0}+\xi({\bf \nabla}-ie{\bf A})]
	\psi_{\sigma}  - \\ \nonumber
	-\Lambda \bar{\psi}_{\uparrow}\bar{\psi}_{\downarrow}
	\psi_{\downarrow}\psi_{\uparrow} + ieA_{0}n_{i}+
	[{\bf E}^{2}+{\bf B}^{2}]/8\pi	{\Big )} \; .
\end{eqnarray}
Here $\xi({\bf \nabla})\equiv -{\bf \nabla}^{2}/2m - \mu$ describes
a single conduction band, and $en_{i}$ denotes the background charge
density of the ions.
The idea is to construct an effective action for the vortex
coordinate ${\bf R}$ only, by integrating out the electronic degrees
of freedom. Our approach is inspired by the one of Simanek
\cite{siman}. In addition to the analysis of Ref.~\cite{siman} we
treat carefully the hydrodynamics of the problem and also avoid
approximations for the matrix elements in the vortex core (see below).

A Hubbard-Stratonovich transformation introduces the energy gap
$\Delta$ as an order parameter field and after performing the trace
over the field $\psi$ (see Refs.~\cite {hkl,aes} for a survey of the
technique used) we arrive at
\begin{eqnarray}
	Z=&\int&{\cal D}^{2}\Delta {\cal D}A_{\alpha} \exp\left(
	\mbox{Tr} \ln {\cal G}^{-1}- S_{0} \right) \;, \\
        \noalign{\vskip 4 pt} \nonumber
	{\cal G}^{-1}&=&\left(\begin{array}{c}
	\partial_{\tau}-ieA_{0}+\xi({\bf \nabla}-ie{\bf A})
	\;\;\;\;\;\;\Delta\\ \noalign{\vskip 2 pt}\bar{\Delta}\;\;\;\;\;\;
	\partial_{\tau}+ieA_{0}-\xi({\bf \nabla}+ie{\bf A})
	\end{array}\right)\;, \\
	S_{0}&=&\int dx\left[\frac{1}{\Lambda}|\Delta|^{2}+
	\frac{{\bf E}^{2}+{\bf B}^{2}}{8\pi}
	+ien_{i}A_{0} \right]\;,  \nonumber
\end{eqnarray}
and $\int dx\equiv\int^{\beta}_{0}d\tau\int d^{3}{\bf r}$. The only
remainder of the electrons is the Nambu-Gor'kov Green's function
${\cal G}$. The Euler-Lagrange equations obtained by varying $A_{0}$
and ${\bf A}$ describe Thomas-Fermi and London screening respectively.
They read \cite{imti}
\begin{eqnarray}\nonumber
	{\bf \nabla}\cdot{\bf E}&=&4\pi i e[n_{e}(\mu+ieA_{0},
		\Delta)-n_{i}]	\; ,\\
	-\partial_{\tau}{\bf E}+ {\bf \nabla}\times
	{\bf B}&=& 4\pi{\bf j}_{e} \; .
\label{maxw}
\end{eqnarray}
Both, the electronic density $n_{e}$ and current density ${\bf j}_{e}$ are
expressed through the electron Green's functions. For
instance, $n_{e}=-\mbox{Tr}[\sigma_{3}{\cal G}]=\frac{1}{2}
\int d\xi N_{e}(\xi+\mu+ieA_0)
[1-\xi/\sqrt{\xi^{2}+\Delta^{2}}]$. The electronic density is a
function of the electro-chemical potential $\mu+ieA_{0}$ and in
the presence of particle-hole asymmetry also of the energy gap
$\Delta$~\cite{fglv,kf}. Due to charge neutrality $n_{e}(\mu,0)=n_{i}$.
Expanding in $A_{0}$ and $\Delta$
we find $n_{e}=n_{i}+ieA_{0}N_{e}+n_{\Delta}+\cdots$, with $n_{\Delta}=
N'_{e}\Delta^{2}/\Lambda N_{e}$. Deviations of $n_{e}$ from $n_{i}$ are
screened on the Thomas-Fermi length $\lambda_{TF}=
(4\pi e^{2}N_{e})^{-1/2}$ and yield a nonzero scalar potential $A_{0}$
determined by the screened Poisson equation
$(-{\bf \nabla}^{2}+\lambda^{-2}_{TF})A_{0}=4\pi ien_{\Delta}$.
Magnetic fields and currents are screened on the scale of the London
penetration depth $\lambda_{L}=(4\pi n_{s}e^{2}/m)^{-1/2}$.
In the following we concentrate on strong Type II superconductors
with $\lambda_{L}\gg\xi$.

Varying $\bar{\Delta}$ yields the BCS gap-equation, which has
a constant as well as vortex solutions.
Here we concentrate on the single vortex solution $\Delta(x)=
\Delta_{v}({\bf r}-{\bf R}(\tau))$ with vortex coordinate
${\bf R}$ and $\Delta_{v}= |\Delta_{v}| e^{i\varphi_{v}}$.
For $\Delta_{v}$ we adopt the mean-field solution from
Ref.~\cite{cgm}. Using $n_{\Delta_{v}}$ as a source, the screened
Poisson equation defines also a single vortex solution $A_{v0}$
for the scalar potential. In the limit of strong screening
($\lambda_{TF}\ll\xi$) $A_{v0}=4\pi ie\lambda^{2}_{TF}n_{\Delta_{v}}$.
As a result, the electronic density in the vortex core does not differ
from the density far away from the core.

We neglect fluctuations around the mean field solutions
$A_{v \alpha}({\bf r}-{\bf R}(\tau))$ and
$\Delta_{v}({\bf r}-{\bf R}(\tau))$, since longitudinal fluctuations
of the phase and $A_{0}$ are lifted to the plasma frequency, transverse
fluctuations of ${\bf A}$ have a gap proportional to the superfluid
density, and fluctuations of $|\Delta|$ are at least at energy
$2\Delta$ \cite{hkl}. Thus, the path-integral measure
$\int{\cal D}^{2}\Delta{\cal D}A_{\alpha}$ reduces to $\int{\cal D}{\bf R}$.

Using a gauge
transformation $eA_{0}\rightarrow eA_{0}-\dot{\varphi}/2\equiv Q_{0}$
and $e{\bf A}\rightarrow e{\bf A}-{\bf \nabla}\varphi/2
\equiv {\bf Q}$, the energy gap $\Delta$ can be chosen real and
manifestly gauge invariant quantities, such as the superfluid
velocity ${\bf Q}/m$, appear in ${\cal G}^{-1}$.
The dynamics of a vortex can now be studied by expanding
$\mbox{Tr}\ln{\cal G}^{-1}$ to second order in $Q_{v\alpha}$
and in the vortex displacement $\delta\Delta=
-{\bf R}(\tau)\cdot{\bf \nabla}\Delta_{v}({\bf r})$ around the static
vortex solution.
Furthermore, due to the singular gauge transformation a source term
${\bf \nabla}\times{\bf \nabla}\varphi_{v}/2e =
\Phi_{0}\delta({\bf r}-{\bf R})$ appears in the London equation that
determines the magnetic field ${\bf B}_{v}$ around a vortex.

We express the unperturbed Nambu-Gor'kov Green's function in the presence
of one vortex in Bogoliubov-de Gennes eigenstates $U_{\lambda}$ with energy
$E_{\lambda}$ as
\begin{equation}
	{\cal G}_{v}({\bf r},{\bf r}';\omega_{\mu})=
	\sum_{\lambda} \frac{U_{\lambda}({\bf r})
	U^{\dagger}_{\lambda}({\bf r}')}
	{i\tilde{\omega}_{\mu}+E_{\lambda}} \; ; \;\;
	U_{\lambda}=\left(\begin{array}{c} u_{\lambda} \\ v_{\lambda}
	\end{array}\right) \;.
\end{equation}
In the relaxation time approximation that we use,
$\tilde{\omega}_{\mu}=\omega_{\mu}+\mbox{sign}(\omega_{\mu})/2\tau_{r}$,
where the $\omega_{\mu}$ are Fermionic Matsubara frequencies.

The result of the expansion is an effective
action $S_{{\rm eff}}[{\bf R}]= S_{C}+S_{H}$ for the coordinate ${\bf R}$
of one vortex, consisting of a hydrodynamic part
\begin{eqnarray}
	S_{H}=\int dx{\Big (}\frac{{\bf E}^{2}_{v}+{\bf B}^{2}_{v}}{8\pi}+
	\frac{1}{2}\Pi_{\alpha\beta}Q_{v \alpha}Q_{v \beta}-
	i n_{\Delta_{v}}Q_{v 0}
	{\Big )} ,\label{sh}
\end{eqnarray}
and a core part
\begin{eqnarray}\nonumber
	S_{C} &=& \frac{1}{2\beta^2}\sum_{\lambda\lambda'}
	\sum_{\mu,n}
	\frac{({\bf R}_{\omega_{n}}\!\!\!\cdot\!\!{\bf W}_{\lambda\lambda'})
	({\bf R}_{-\omega_{n}}\!\!\!\cdot\!\!{\bf W}_{\lambda'\lambda})}
	{(i\tilde{\omega}_{\mu}+E_{\lambda})
	(i\tilde{\omega}_{\mu +n}+E_{\lambda'})} \, ,\\
        \noalign{\vskip 5 pt}
	{\bf W}_{\lambda\lambda'}&=&\int d^{2}r U^{\dagger}_{\lambda}
		\left[\begin{array}{cc} 0 &{\bf \nabla}\Delta_{v} \\
		{\bf \nabla}\bar{\Delta}_{v} & 0 \end{array}\right]
	U_{\lambda'}  \; ,\label{sc}
\end{eqnarray}
and the $\omega_{n}$ denote Bosonic Matsubara frequencies.

First we discuss {\it the hydrodynamic contribution} $S_{H}$.
The kernel $\Pi_{\alpha\beta}$ is the polarization bubble and
describes both longitudinal and transverse screening \cite{larmig}.
The transverse part of the polarization term in Eq.~(\ref{sh})
is ${\bf \Pi}_{T}=n_{s}(T)(\delta^{ab}-k^{a}k^{b}/k^{2})+\cdots$,
where $a,b=x,y,z $ and the $\cdots$ indicate higher order terms
in ${\bf k}$ and $\omega$. The longitudinal part of the polarization
term in Eq.~(\ref{sh}) is $\Pi_{L}=N_{e}+\cdots$.
Using the equation of motion (\ref{maxw}) for the fields and
$n_{e}(\mu,\Delta_{v})-n_{i}= n_{\Delta_{v}}$, the hydrodynamic part
of the effective action can be rewritten as
\begin{eqnarray}\nonumber
	S_{H}=\frac{1}{8\pi}\int dx {\Big(}  {\bf E}^{2}_{v}+
	{\bf B}^{2}_{v}+\lambda^{2}_{L}
	(-\partial_\tau {\bf E}_v + {\bf \nabla}\times{\bf B}_{v})^{2} + \\
	\lambda^{2}_{TF}({\bf \nabla}\cdot{\bf E}_{v})^{2} +
	16\pi^{2}\lambda^{2}_{TF} n^{2}_{\Delta_{v}}
	+4\pi i n_{\Delta_{v}} {\dot\varphi}_{v}    {\Big)} \; .
\label{hydr}
\end{eqnarray}
The last term is the most important one for our discussion as it yields
the hydrodynamic contribution to the Hall coefficient $\gamma_{H}=
\pi n_{\Delta}=\pi N_{e}'\Delta^{2}/(\Lambda N_{e})$. This result
coincides with that of the TDGL-approach. The only other dynamic term
in Eq.~(\ref{hydr}) is the transverse part of the ${\bf E}^{2}$ term
that yields a small electromagnetic mass~\cite{suhl,futur}.
All other terms are non-dynamic and contribute to the line energy of
a vortex \cite{rmp}.

We now turn to {\it the core contribution} $S_{C}$ in Eq.~(\ref{sc}).
Its origin is found in the transitions induced by the moving vortex
between the CdGM states in the core labeled by $\lambda$ and
$\lambda'$ and involving the matrix elements $W_{\lambda\lambda'}$.
The energies are $E_{\lambda}=\lambda\omega_{0}$ with $\lambda=$
half-integer. The sums over states $\lambda$ and
$\lambda'$ in Eq.~(\ref{sc}) may be evaluated using the constant
level separation $E_{\lambda}-E_{\lambda-1}=\omega_{0}$ \cite{cgm}
and properties of the matrix elements \cite{kks}. Explicitly we use
the identity
\begin{equation}
	U^{\dagger}_{\lambda}
                \left[\begin{array}{cc} 0 &{\bf \nabla}\Delta_{v} \\
                {\bf \nabla}\bar{\Delta}_{v} & 0 \end{array}\right]
        U_{\lambda'}
	= (E_{\lambda'}-E_{\lambda}) U^{\dagger}_{\lambda}
	{\bf \nabla}U_{\lambda'} \; ,
\end{equation}
together with the relations for the eigenstate wavefunctions
$\nabla_{x}U_{\lambda}= (k_{F}/2)[U_{\lambda-1}-U_{\lambda+1}]$,
$\nabla_{y}U_{\lambda}= (ik_{F}/2)[U_{\lambda-1}+U_{\lambda+1}]$,
see Ref.~\cite{kks} for a discussion of this point. Using the
orthogonality relations, we find the selection rule that only
neighboring states are connected by the matrix elements.
We also restrict ourselves to the temperature range $\omega_0 \ll T$
where the sum over $\lambda$'s may be replaced by the integral
$\int dE_{\lambda}/\omega_0$.
The result for $S_{C}$ can be written in the form
\begin{eqnarray}\nonumber
	S_{C}=\frac{1}{2}\int^{\beta}_{0} d\tau\int^{\beta}_{0} d\tau'
	{\Big [}
	K^{+}_{C}(\tau-\tau')\, {\bf R}(\tau)\cdot{\bf R}(\tau')+\\
	+i K^{-}_{C}(\tau-\tau')\, {\bf z}\cdot({\bf R}(\tau)
	\times{\bf R}(\tau')) {\Big ]} \; .
\label{core}
\end{eqnarray}
In Fourier components the kernels $K^{\pm}_{C}$ determining
the mass and damping ($K^{+}$) and Hall force ($K^{-}$) are
\begin{equation}
	K^{\pm}_{C}(\omega_{n})=\frac{\omega_{0}k^{2}_{F}}{4}
	\left[\frac{i\omega_{n}}{i\bar{\omega}_{n}-\omega_{0}}
	\pm\frac{i\omega_{n}}{i\bar{\omega}_{n}+\omega_{0}}\right] \; ,
\end{equation}
where $\bar{\omega}_{n}=\omega_{n}+\tau^{-1}_{r}\mbox{sign}(\omega_{n})$.
They are non-local in time, however, after analytic continuation to real
frequencies they can be expanded in $\omega/\omega_{0}$.
The kernel $K^{-}_{C} \approx -i\omega_{n}
(k^{2}_{F}\omega^{2}_{0}\tau^{2}_{r}/2)/
[1+(\omega_{0}\tau_{r})^{2}]$, yields the
core contribution $\gamma_{C}$ as quoted in Eq.~(\ref{class}),
if we put $k_{F}^{2}=2\pi n_{e}$ in two dimensions.
The kernel $K^{+}_{C} \approx (|\omega_{n}|+\tau_{r}\omega^{2}_{n})
(k^{2}_{F}\omega_{0}\tau_{r}/2)/[1+(\omega_{0}\tau_{r})^{2}]$
is proportional to $|\omega_{n}|$ for small frequencies, thus
describing Ohmic dissipation \cite{cl,aes}. Apart from the damping
coefficient $\eta_{C}$ it yields the core contribution to the
vortex mass \cite{kks,siman},
\begin{equation}
	M_{C}=\frac{(\omega_{0}\tau_{r})^{2}}{[1+(\omega_{0}\tau_{r})^{2}]}
	\left(\frac{\epsilon_{F}}{\Delta}\right)^{2}m \; ,
\end{equation}
which is large in the superclean limit with $\omega_{0}\tau_{r}\gg 1$.

Thus, a complete description of intrinsic vortex properties
can be obtained if both core and
hydrodynamic contributions are included. The hydrodynamic part of
the Hall-force was neglected in Refs.~\cite{kks,koplop,siman}, whereas
the core physics cannot be described by a hydrodynamic theory
such as TDGL \cite{tdgl}.
In Ref.~\cite{aoth} the superconducting phase was coupled to the
superfluid density in order to obtain a Galilei-invariant
Magnus force ${\bf F}_{M}= \kappa({\bf V}_{T}-{\bf V})\times{\bf z}$
from hydrodynamics only. Our analysis shows that the phase of the
superconducting order parameter couples to the square of
the order parameter, with a small coefficient that depends on
particle-hole asymmetry, {\it i.~e.}, details of the electronic band
structure are relevant. A Galilei invariant Magnus force at $T=0$ and
$\tau_{r}=\infty$ in Fermionic superconductors is provided by the
vortex core rather than the hydrodynamic flow around the vortex.

In conclusion we have presented a microscopic derivation of the
equation of motion of a vortex in a superconductor. Our results
relate the observed sign change in the Hall effect in
superconductors with $l\sim\xi$ to broken particle-hole symmetry
in the electronic band structure.

We thank U. Eckern, R. Fazio, E. Heeb,
R.P. H\"{u}bener, N.B. Kopnin, A.I. Larkin, G. Sch\"{o}n, M. Skvortsov,
and G.T. Zimanyi for discussion, and E. Simanek for sending
a preprint of his work before publication. One of us (AvO) thanks
the Landau Institute, where part of this work was done, for
hospitality. A joint grant from the International Science Foundation
and the Russian Government (M6M300), a grant from the Russian Foundation for
Fundamental Research (\# 95-02-05720), and support by the Swiss National
Foundation are gratefully acknowledged.


\begin{references}
\bibitem{imti}  In the imaginary time formulation $i=\sqrt{-1}$ appears.
		To make contact to the real time formulation replace
		$\tau\rightarrow it$ and $A_{0}\rightarrow -iA_{0}$.
\end{references}
\end{document}